\documentclass[%
reprint,
nofootinbib,
amsmath,amssymb,
aps,
]{revtex4-2}

\usepackage{graphicx}
\usepackage{amssymb}
\usepackage{amsmath}
\usepackage{mathbbol}
\usepackage{amsthm}
\usepackage[normalem]{ulem}
\usepackage{slashed}
\usepackage{color,rotating}
\usepackage{bbm}
\usepackage{array}
\usepackage[utf8]{inputenc}
\usepackage{multirow}
\usepackage{xcolor}

\allowdisplaybreaks
\usepackage[toc,page]{appendix}

\newcommand{\p}{\partial}

\newcommand{\gb}{\bar{g}}

\newcommand{\cG}{\mathcal{G}}
\newcommand{\cM}{\mathcal{M}}

\newcommand{\cR}{\mathcal{R}}

\newcommand{\be}{\begin{equation}}
\newcommand{\ee}{\end{equation}}

\begin{document}

\title{Global Flows of Foliated Gravity-Matter Systems}

\author{Guus Korver}
\email{guus.korver@ru.nl}

\author{Frank Saueressig}
\email{f.saueressig@science.ru.nl}

\author{Jian Wang}
\email{Jian.Wang@science.ru.nl}

	\affiliation{Institute for Mathematics, Astrophysics and Particle Physics (IMAPP) \\ Radboud University, Heyendaalseweg 135, 6525 AJ Nijmegen,The Netherlands 
}

\begin{abstract}
Asymptotic safety is a promising mechanism for obtaining a consistent and predictive quantum theory for gravity. The ADM formalism allows to introduce a (Euclidean) time-direction in this framework. It equips spacetime with a foliation structure by encoding the gravitational degrees of freedom in a lapse function, shift vector, and a metric measuring distances on the spatial slices. We use the Wetterich equation to study the renormalization group flow of the graviton $2$-point function extracted from the spatial metric. The flow is driven by the $3$- and $4$-point vertices generated by the foliated Einstein-Hilbert action supplemented by minimally coupled scalar and vector fields. We derive bounds on the number of matter fields cast by asymptotic safety. Moreover, we show that the phase diagram obtained in the pure gravity case is qualitatively stable within these bounds. An intriguing feature is the presence of an IR-fixed point for the graviton mass which prevents the squared mass taking negative values. This feature persists for any number of matter fields and, in particular, also in situations where there is no suitable interacting fixed point rendering the theory asymptotically safe. Our work complements earlier studies of the subject by taking contributions from the matter fields into account.  
\end{abstract}

\maketitle

%\begin{keyword}
%Functional Renormalization Group \sep Asymptotic Safety \sep Gravity-Matter Systems \sep 

%\PACS{04.60.-m,04.62.+v,11.10.Hi} 

%\end{keyword}

%\end{frontmatter}

%-------------------------------------------------------
%\include{ADM}
%\include{FRGE}
%\include{RG flow pure gravity}
%-------------------------------------------------------
\section{Introduction}
%-------------------------------------------------------
General relativity describes the working of gravity in terms of a classical field theory while all other known fundamental forces are modeled in terms of a relativistic quantum field theory. Bringing these constructions to an equal theoretical footing then calls for a quantum theory for gravity. Conservatively, one may expect that such a theory may modify the laws of gravity around the Planck scale $M_{\rm Pl} \approx 10^{19}$ GeV. Classical general relativity then acquires the status of an effective field theory capturing the gravitational dynamics at low energy \cite{Donoghue:2017pgk}.

The vast separation between the Planck scale and scales accessible experimentally poses a significant challenge when seeking to connect proposals for quantum gravity to observations. There is one benchmark, which may be non-trivial from the quantum gravity perspective: the propagator capturing gravitational fluctuations in a flat spacetime should be massless. This feature, inherent to general relativity, is closely linked to cosmological perturbation theory and also the propagation of gravitational waves \cite{Addazi:2021xuf}.

From the perspective of the quantum theory, one may investigate this condition as follows. Starting from a metric $g_{\mu\nu}$ on a four-dimensional (Euclidean) spacetime, one may use the Arnowitt-Deser-Misner (ADM) decomposition \cite{Arnowitt:1962hi,Gourgoulhon:2007ue} in order to encode the gravitational degrees of freedom in terms of a lapse function $N$, a shift vector $N_i$ and a metric $\sigma_{ij}$ measuring distances on the hypersurfaces of constant time. 

Subsequently, one can consider fluctuations in $\sigma_{ij}$ around a flat background. Extracting the transverse-traceless part of these fluctuations, denoted by $h_{ij}$ in the sequel, then gives the components typically associated with gravitational waves. Restricting to two spacetime-derivatives and insisting on locality, the part of the (quantum) effective action describing the propagation of these degrees of freedom in this situation has the form
\be\label{Gammaeffans}
\Gamma = \frac{1}{64 \pi G} \int d^4x \, h_{ij} \left[ -\p^2 + \mu^2 \right] \, h^{ij} \, . 
\ee 
Here $G$ may be read as a constant which can be absorbed in the normalization of the fluctuation field and $\mu^2$ is the renormalized squared mass. The quantum theory should then yield $\mu^2 = 0$ in order to agree with observations.

The Wetterich equation \cite{Wetterich:1992yh,Morris1994,Reuter:1993kw}, adapted to gravity in \cite{Reuter:1996cp}, provides a powerful tool for computing the admissible values of $\mu^2$. Technically, this approach implements Wilson's modern viewpoint on renormalization, integrating out quantum fluctuations shell-by-shell in momentum space. Thus it applies to both an effective field theory setting as well as to the case where there is a high-energy completion of the theory by a suitable renormalization group (RG) fixed point along the lines of the gravitational asymptotic safety program recently summarized in the Handbook of Quantum Gravity \cite{Saueressig:2023irs,Pawlowski:2023gym,Eichhorn:2022gku,Knorr:2022dsx,Morris:2022btf,Martini:2022sll,Wetterich:2022ncl,Platania:2023srt}, the pedagogical expositions \cite{Codello:2008vh,Nagy:2012ef,Reichert:2020mja}, and textbooks \cite{Percacci:2017fkn,Reuter:2019byg}.

The Wetterich equation adapted to the ADM formalism has been constructed in \cite{Manrique:2011jc,Rechenberger:2012dt}. Evaluating the resulting RG at zeroth order in the fluctuation fields \cite{Biemans:2016rvp,Biemans:2017zca,Houthoff:2017oam} indicated that the formulation gives rise to a RG fixed point with similar properties as found in the covariant formulation \cite{Reuter:1996cp,Reuter:2001ag,Lauscher:2001ya,Litim:2003vp,Gies:2015tca}. In order to track the RG flow of the $2$-point correlator \eqref{Gammaeffans}, one has to go beyond the background approximation and resort to the fluctuation approach \cite{Christiansen:2012rx,Christiansen:2014raa,Christiansen:2015rva,Denz:2016qks,Christiansen:2017cxa,Christiansen:2017bsy,Eichhorn:2018akn}, reviewed in \cite{Pawlowski:2020qer,Pawlowski:2023gym}. In the covariant setting, this approach has recently provided significant insights on the momentum-dependence of the graviton $2$-point function \cite{Knorr:2021niv,Bonanno:2021squ,Fehre:2021eob}.
 Fluctuation computations then allow to track the change of the couplings $G, \mu^2$ in \eqref{Gammaeffans} when integrating out quantum fluctuations also in the ADM formalism. For the pure gravity theory this strategy has recently been implemented in \cite{Saueressig:2023tfy}. As the main results, it has been shown that $G$ and $\mu^2$ admit a non-Gaussian fixed point (NGFP) suitable for rendering the theory asymptotically safe. Moreover, the phase space analysis revealed the presence of an IR fixed point (IR-FP) which drives the squared mass to zero dynamically. A similar IR-FP also appears in the covariant construction \cite{Christiansen:2014raa}.

The present work extends these pure-gravity results \cite{Saueressig:2023tfy} by including matter degrees of freedom in the form of $N_s$ scalar fields and $N_v$ abelian gauge fields which are minimally coupled to gravity. This leads to two remarkable insights. Firstly, the NGFP which renders the pure gravity theory asymptotically safe
admits continuous deformations depending on the number of matter fields. We then derive bounds on $N_s$ and $N_v$ for which this family of fixed points persists. Secondly, we find that the IR-FP appearing in the low-energy regime of the phase diagram persists for any number of matter fields. In particular, the mechanism is independent of whether the RG flow admits a high-energy completion or one is dealing with an effective field theory. Thus, the mechanism may be operative in any theory of quantum gravity which admits an effective field theory description at low energy scales.

The rest of our work is organized as follows. In Sect.\ \ref{sect.2} we start by reviewing the ADM decomposition and the Wetterich equation adapted to this framework. Sect.\ \ref{sect.3} describes our projection of the RG flow and gives the resulting beta functions. The resulting fixed point structure and phase diagrams are constructed in Sect.\ \ref{sect.4} and we conclude with our conclusions and a brief outlook in Sect.\ \ref{sect.conclusions}.

%-------------------------------------------------------
\section{Background and Computational Framework}
\label{sect.2}
%-------------------------------------------------------
We start by introducing the Wetterich equation for a foliated spacetime. The discussion is kept minimalistic and we refer to \cite{Rechenberger:2012dt,Saueressig:2023tfy} for more details. 

We start from a $d+1$-dimensional spacetime $\mathcal{M}$, equipped with coordinates $x^{\mu}$, $\mu\in\{0,...,d\}$ and a Euclidean metric $g_{\mu\nu}$. On this spacetime, we introduce a scalar function $\tau: \mathcal{M}\rightarrow\mathbb{R}$ that defines the hypersurfaces $\Sigma_{\tau_i}:=\{x\in \mathcal{M}|\tau(x)=\tau_i\}$. The hypersurfaces equip $\cM$ with a foliation structure so that $\cM = \mathbb{R}\times\Sigma$. Following the geometrical construction reviewed e.g.\ in \cite{Gourgoulhon:2007ue}, we can introduce a new set of coordinates adapted to the foliation structure $x^\mu \mapsto (\tau, y^i)$, $i \in \{1,\cdots,d\}$. In these new coordinates the line-element then takes the form
\be\label{dsfoliated}
ds^2 = N^2 d\tau^2 + \sigma_{ij} \left(N^i d\tau + dy^i\right)  \left(N^j d\tau + dy^j\right) \, . 
\ee
The lapse function $N$, shift vector $N_i$ and the metric $\sigma_{ij}$ on the hypersurfaces $\Sigma_\tau$ constitute the ADM fields. Their relation to the spacetime metric is given by
\be\label{ADMdecomposition}
\begin{split}
 \left(g_{\mu\nu}\right)=  & 
\begin{bmatrix}
	N^2+N^i N_i \quad& N_j \\
	N_i \quad& \sigma_{ij} \\
\end{bmatrix} \, , \; \\
 \left(g^{\mu\nu}\right)= & 
\begin{bmatrix}
	\frac{1}{N^2} \quad& -\frac{N^j}{N^2} \\
	-\frac{N^i}{N^2} \quad& \sigma^{ij}+\frac{N^i N^j}{N^2} 
\end{bmatrix} \, . 
\end{split}
\ee
Notably, the relation between $g_{\mu\nu}$ and the ADM fields is non-linear. Moreover, the analytic continuation of the lapse function $N \mapsto i N$ allows to transit from the Euclidean to the Lorentzian setting without the line-element \eqref{dsfoliated} becoming complex, also see \cite{Baldazzi:2018mtl} for a recent and more detailed discussion.

The Wetterich equation on foliated spacetimes \cite{Manrique:2011jc} encodes the gravitational degrees of freedom in the ADM fields. In order to be able to introduce a reference scale which allows to separate quantum fluctuations into high- and low-momentum modes, the ADM fields are decomposed into fixed background values and fluctuations about this background
\begin{equation}\label{backgrounddecomposition}
	\sigma_{ij}=\Bar{\sigma}_{ij}+\hat{\sigma}_{ij},\quad N_i=\Bar{N}_i+\hat{N}_i,\quad N=\Bar{N}+\hat{N}.
\end{equation}
Conceptually, this corresponds to quantizing a theory along the lines of a quantum field theory in curved spacetime set by the background fields $(\Bar{\sigma}_{ij}, \Bar{N}_i, \Bar{N})$. In general, one could keep the background unspecified, quantizing the fields $(\hat{\sigma}_{ij},\hat{N}_i,\hat{N})$ in all backgrounds simultaneously. This underlies the background independence of the construction based on the philosophy that there is no preferred background structure to work with \cite{Becker:2014qya}. In this work, we will be less ambitious though and choose the background metric as a flat, four-dimensional Euclidean spacetime
\begin{equation}\label{choicebackground}
	\Bar{\sigma}_{ij}=\delta_{ij}, \quad \Bar{N}=1, \quad \Bar{N}_i=0.
\end{equation}
This choice has the virtue that quantum corrections can be computed by resorting to standard momentum space techniques which leads to drastic simplifications at the technical level. In order to further simplify the computation and getting access to the correlation function \eqref{Gammaeffans}, we furthermore perform a York decomposition \cite{York1973} of the fluctuation fields.\footnote{The component fields appearing in this way are also closely related to the fields used in cosmic perturbation theory. In particular $h_{ij}$ encodes the tensor fluctuations in a gauge-invariant way \cite{Bardeen:1980kt,Craps:2014wga}.} Defining the background Laplacian $\Delta \equiv - \delta^{ij} \p_i \p_j$, the shift vector is separated into its transverse and longitudinal component
\begin{equation}
	\label{eqn: York decomp shift}
	\hat{N_i}=u_i+\partial_i \frac{1}{\sqrt{\Delta}}B, \quad \partial^i u_i=0 \, . 
\end{equation}
Along similar lines, $\hat{\sigma}_{ij}$ is written in terms of a transverse-traceless (TT)-tensor mode $h_{ij}$, a transverse vector mode $v_i$, a scalar mode $E$, and a trace mode $\psi \equiv \delta^{ij}\hat{\sigma}_{ij}$,
\begin{equation}
	\label{eqn: York decomp metric}
	\hat{\sigma}_{ij}=h_{ij}+\partial_i \frac{1}{\sqrt{\Delta}}v_j+\partial_j \frac{1}{\sqrt{\Delta}}v_i +\partial_i \partial_j \frac{1}{\Delta}E+\frac{1}{3}\delta_{ij} E+\frac{1}{3}\delta_{ij}\psi \, , 
\end{equation}
where the component fields satisfy
\begin{equation}
	\partial^i h_{ij}=0, \quad \delta^{ij}h_{ij}=0, \quad \partial^i v_i=0 \, . 
\end{equation}
Here the inverse powers of the Laplacian ensure that all component fields have the same mass dimension. They also ensure the absence of Jacobians arising from the field redefinition. At the technical level, the decomposition further simplifies the momentum integrations since all contractions of a partial derivative with a field vanish due to the transversality constraints \cite{Benedetti:2010nr}. 

In order to study the impact of quantum fluctuations on the $2$-point function \eqref{Gammaeffans}, we use the Wetterich equation \eqref{eqn:Wetterich}. This equation describes the change of the effective average action $\Gamma_k$ when integrating out quantum fluctuations with momenta $p^2$ comparable to the coarse-graining scale $k$. Its derivation follows the standard lines \cite{Codello:2008vh,Percacci:2017fkn,Reuter:2019byg,Saueressig:2023irs}, and results in the standard form
\begin{equation}
	\label{eqn:Wetterich}
	\partial_t \Gamma_k = \frac{1}{2} \mathrm{STr} \left[ \left(\Gamma^{(2)}_k+\mathcal{R}_k \right)^{-1} \,  \partial_t \mathcal{R}_k\right] \, . 
\end{equation}
Here $t \equiv \ln(k/k_0)$ is the RG time with $k_0$ being an arbitrary reference scale. Furthermore, $\Gamma^{(2)}_k$ denotes the second functional derivative of $\Gamma_k$ with respect to the fluctuation fields, which is matrix-valued in field space. The supertrace STr on the right-hand side then contains an integration over loop-momenta, a sum over all fluctuation fields and a minus sign for ghost contributions.  The regulator $\mathcal{R}_k$ equips fluctuations with momenta $p^2 \lesssim k^2$ with a $k$-dependent mass term and vanishes for high-momentum modes $p^2 \gg k^2$. Moreover, the condition that $\cR_k \propto k^2$ ensures that in the limit $k^2 \rightarrow 0$ all fluctuations are integrated out. The interplay of the regulator terms appearing on the right-hand side ensures that the argument vanishes for momenta $p^2 \gg k^2$ and is peaked around the coarse-graining scale. As a consequence, the trace comes with a built-in UV-regulator and the flow of $\Gamma_k$ is driven by integrating out quantum fluctuations with momenta $p^2\gtrsim k^2$. In this sense, the Wetterich equation implements a Wilsonian viewpoint on renormalization  \cite{Wilson:1973jj}, integrating out quantum fluctuations shell-by-shell in momentum space. In particular, one recovers the standard effective action $\Gamma$ in the limit $\lim_{k\rightarrow 0} \Gamma_k = \Gamma$, provided that this limit exists. 

A key feature of the Wetterich equation is that the construction of its solutions does not require specifying a fundamental action beforehand. One may simply impose initial conditions $\Gamma_{\Lambda_{\rm UV}}$ at $\Lambda_{\rm UV}$ and construct the resulting effective action by integrating the flow equation to $k=0$. In this way, the flow equation allows to compute quantum corrections in the framework of effective field theories, independent of whether $\Gamma_{\Lambda_{UV}}$ is derived from a high-energy completion.

A special role is then played by solutions $k \mapsto \Gamma_k$ which are complete in the sense that they are well-defined for all values of the coarse-graining scale $k^2 \in [0, \infty[$. One way to ensure the existence of a well-defined limit $k \rightarrow \infty$ is by demanding that the solution approaches a RG fixed point in this limit. In this case the dimensionless couplings in $\Gamma_k$ remain finite and physical quantities are free from unphysical UV-divergences. In the context of quantum gravity a central role is played by NGFP corresponding to interacting quantum field theories. These are the key ingredient for providing a well-defined and predictive high-energy completion of gravity and gravity-matter systems along the lines of the gravitational asymptotic safety program \cite{Saueressig:2023irs,Pawlowski:2023gym,Knorr:2022dsx,Morris:2022btf}. In this context the Wetterich equation has yielded substantial evidence that gravity indeed possesses a suitable RG fixed point (the Reuter fixed point) for rendering the theory asymptotically safe \cite{Lauscher:2002sq,Codello:2007bd,Machado:2007ea,Benedetti:2009rx,Gies:2016con,Falls:2017lst,Falls:2018ylp,Knorr:2021slg,Baldazzi:2023pep}. Qualitatively similar fixed points (deformed Reuter fixed points) exist for a wide range of gravity-matter systems \cite{Eichhorn:2018yfc,Eichhorn:2022gku}. In particular, it is likely that gravity supplemented by the standard model degrees of freedom admits a UV-completion via the asymptotic safety mechanism \cite{Dona:2013qba,Christiansen:2017cxa,Alkofer:2018fxj,Eichhorn:2018ydy}, see, in particular \cite{Pastor-Gutierrez:2022nki} for recent results and more references.

%-------------------------------------------------------
\section{Projecting of the RG flow and beta functions}
\label{sect.3}
%-------------------------------------------------------
In this project, we are interested in the \emph{Wilsonian renormalization group flow} of \eqref{Gammaeffans} in the presence of scalar and gauge fields. To this end, we promote the two couplings $G$ and $\mu$ to functions of the coarse graining scale $k$. The resulting $\Gamma_k$ is then taken to span the space on which the RG flow encoded in the Wetterich equation \eqref{eqn:Wetterich} is projected. 

At this point, we note that the action \eqref{Gammaeffans} is of second order in the fluctuation field $h_{ij}$. Following the idea of the vertex expansion \cite{Pawlowski:2020qer,Pawlowski:2023gym}, we obtain the equation encoding the RG flow of this function by taking functional derivatives of \eqref{eqn:Wetterich}. Denoting the vertices involving $n$ fluctuation fields by $\Gamma_k^{(n)}$, and abbreviating the regulated propagator as $\cG_k \equiv \left(\Gamma_k^{(2)} + \cR_k\right)^{-1}$, the resulting flow equation for the $2$-point function takes the schematic form
\be\label{eqn:flow:2pt}
\begin{split}
\p_t \Gamma_k^{(2)} = & \, {\rm STr}\Big[ \cG_k  \; \Gamma_k^{(3)} \; \cG_k   \Gamma_k^{(3)} \; \cG_k \;  \p_t \cR_k \Big] \\
& - \frac{1}{2}{\rm STr}\left[ \cG_k  \; \Gamma_k^{(4)} \; \cG_k  \; \p_t \cR_k \right] \, . 
\end{split}
\ee
The ``external'' fluctuation fields on the right-hand side are fixed to be $h_{ij}$. This entails that the $3$-point and $4$-point vertices on the right-hand side must come with one (respectively two) ``external legs'' associated with $h_{ij}$ in order to match the structure on the left-hand side. The STr indicates that all fluctuation fields propagate in the loop though.

Eq.\ \eqref{eqn:flow:2pt} shows that the flow of the $2$-point function is driven by $2$-, $3$-, and $4$-point vertices. This information is not contained in eq.\ \eqref{Gammaeffans}. Thus, we must provide a closure condition for the projected equation. This can, e.g., be done along the BMW-scheme \cite{Blaizot:2005xy,Benitez:2011xx}. Here, we follow a slightly different route and generate all propagators and vertices from the Einstein-Hilbert (EH) action supplemented by minimally coupled matter fields and supplemented by suitable gauge-fixing (gf) and ghost terms
\be\label{Gansscheme}
\Gamma_k \simeq \Gamma_k^{\rm EH} + \Gamma_k^{\rm gf} + S^{\rm ghost} + S^{\rm scalar} + S^{\rm vector} \, . 
\ee 
Here we use the symbols $\Gamma_k$ and $S$ to indicate whether the corresponding actions contain $k$-dependent couplings and we use the $\simeq$-symbol in order to indicate that we are working with an approximation of the full effective average action. The EH-action written in terms of the ADM fields is given by
\begin{equation}\label{eqn:EH-ADM}
	\begin{split}
	\Gamma^{\rm EH}_k = \frac{1}{16 \pi G_k} & \int \mathrm{d}\tau \mathrm{d}^3 y N\sqrt{\sigma} \times \\ & \times\left(K^{ij}K_{ij}-K^2-{}^{(3)}\hspace{-0.2mm}R+2\Lambda_k \right). 
	\end{split}
\end{equation}
It contains the scale-dependent Newton's coupling $G_k$ and the cosmological constant $\Lambda_k$. The extrinsic curvature tensor $K_{ij}$ is defined as
\begin{equation}
	K_{ij} \equiv \frac{1}{2N}\left(\partial_{\tau}\sigma_{ij}-D_i N_j-D_j N_i \right) \, , 
\end{equation}
where $D_i$ denotes the spatial covariant constructed from $\sigma_{ij}$. Furthermore, 
 $K \equiv \sigma^{ij}K_{ij}$ and $^{(3)}R$ is the Ricci scalar obtained from $\sigma_{ij}$, encoding the intrinsic curvature of the spatial hypersurfaces. Following \cite{Biemans:2016rvp,Houthoff:2017oam}, we supplement the EH-action by the gauge-fixing action 
 \begin{equation}
 	\label{eqn:gfaction}
 	\Gamma^{\text{gf}}_k=\frac{1}{32 \pi G_k}\int \mathrm{d}\tau \mathrm{d}^3 y \left(F^2 +\delta^{ij}F_{i} F_{j}\right) \, . 
 \end{equation}
 The gauge-fixing conditions $F$ and $F_i$ are linear in the fluctuations fields  
 \be\label{eqn:gffunctionals}
 \begin{split}
 	F &= \partial_{\tau}\hat{N}+  \partial^i \hat{N}_i - \frac{1}{2} \partial_{\tau} \psi \, , \\
 	F_i &=  \partial_{\tau}\hat{N}_i - \partial_i \hat{N} - \frac{1}{2} \partial_i  \psi +  \partial^j \hat{\sigma}_{ij} . 
 \end{split}
 \ee
 The ghost action accompanying this gauge-fixing functional contains a scalar and vector (anti-)ghost pair $(\bar{c},c,\bar{b}^i,b_i)$. The explicit expression is rather lengthy and has been relegated to App.\ \ref{App.ghosts}. The gauge-fixing condition is distinguished by the property that it leads to relativistic $2$-point functions for all component fields.
 
 The action for the $N_s$ minimally coupled scalar fields $\phi$ is
 \be\label{eq:Sscalar}
 S^{\text{scalar}} = \frac{1}{2}\sum_{n=1}^{N_s} \int \mathrm{d}\tau \mathrm{d}^3 y N \sqrt{\sigma} \, g^{\mu\nu} \, \big(\p_\mu \phi \big) \big(\p_\nu \phi \big) \, , 
 \ee 
 where $g^{\mu\nu}$ should be read as a short-hand notation for the ADM fields via the substitution \eqref{ADMdecomposition}. In the vector sector, we use
 \be\label{eq:Svector}
 \begin{split}
 		S^{\text{vector}} = & \, \frac{1}{4} \sum_{n=1}^{N_v}\int \mathrm{d}\tau \mathrm{d}^3 y \,  N\sqrt{\sigma} \, g^{\mu\alpha} g^{\nu\beta} F_{\mu\nu} F_{\alpha\beta} \\
 		& \, + \frac{1}{2} \sum_{n=1}^{N_v} \int \mathrm{d}\tau \mathrm{d}^3 y \, \big(\p^\mu A_\mu \big) \big(\p^\nu A_\nu \big) \\
 		& + \sum_{n=1}^{N_v}\int \mathrm{d}\tau \mathrm{d}^3 y \, \bar{C} \left[-\p^2 \right] C \, , 
 \end{split}
 \ee
 where $F_{\mu\nu} \equiv \p_\mu A_\nu - \p_\nu A_\mu$ is the field strength tensor. The gauge-fixing is constructed using the background metric. This suffices in obtaining a well-defined propagator. The second and third line do not give rise to $3$- and $4$-point vertices. Hence the matter ghost fields $\bar{C}, C$ do not contribute to the right-hand side of the flow equation. Notably, it is equally plausible to construct the gauge-fixing and ghost terms in eq.\ \eqref{eq:Svector} using the full metric $g_{\mu\nu}$. This creates additional vertices containing external legs associated with $h_{ij}$. We will not follow this option in the sequel and adopt eqs.\ \eqref{eqn:EH-ADM}, \eqref{eqn:gffunctionals}, \eqref{Sghost}, \eqref{Vghost}, \eqref{eq:Sscalar}, and \eqref{eq:Svector} as our closure of the projected RG equation.

The $2$-, $3$-, and $4$-point vertices featuring in the computation are obtained by substituting the linear split \eqref{backgrounddecomposition}, specifying the background according to \eqref{choicebackground}, and expanding in the fluctuation fields. The background value of the ghost- and matter fields are set to zero for computational simplicity. The explicit form of the propagators and vertices is readily obtained via the xAct-package \cite{xactref,Nutma:2013zea} for {\tt Mathematica}. In particular, the expansion recovers the $2$-point function \eqref{Gammaeffans} upon identifying 
\be
\mu_k^2 \equiv -2 \Lambda_k \, .
\ee
 The remaining propagators and vertices depend on $G_k$ and $\Lambda_k$ only. Substituting them into \eqref{eqn:flow:2pt} then yields a closed expression for the Wilsonian RG flow of these couplings.

Throughout this work, we use a regulator function $\cR_k$ which implements the idea of having a $k$-dependent mass term for all fluctuation modes through the rule $p^2 \mapsto p^2 + R_k(p^2)$ (Type I regulator in the notation of \cite{Codello:2008vh}). The scalar regulator is taken to be of Litim-type \cite{Litim:2000ci,Litim:2001up}, 
\be\label{Rprofile}
R_k(p^2) = (k^2-p^2) \Theta(k^2-p^2) \, , 
\ee
where $\Theta(x)$ is the Heaviside step function. Following the pure gravity computation \cite{Saueressig:2023tfy}, the traces appearing on the right-hand side can now be evaluated using standard Feynman diagram techniques. The regulator ensures that all loop integrals are finite and can be evaluated analytically.

At this stage the following technical comment on the projection is in order. The $2$-point function \eqref{Gammaeffans} depends on the $4$-momentum $p^2 = p_0^2 + \vec{p}^{\,2}$ where $p_0$ and $\vec{p}$ denote the ``time'' and spatial component of the momentum 4-vector, respectively. Since the foliation allows to distinguish between these two entries, the flow of $G_k$ can be read off from either the terms proportional to $p_0^2$ or $\vec{p}^{\,2}$. In case of a fully relativistic theory, their results will agree due to Lorentz symmetry. A careful analysis of the vertices generated in the expansion procedure shows, however, that the $3$- and $4$-point vertices including the fluctuation of the lapse function $\hat{N}$ carry a dependence on $p_0^2$ and $\vec{p}^{\,2}$ which does not combine into a relativistic momentum $4$-vector.\footnote{Qualitatively, one can convince oneself about this feature by observing that the gauge-fixing term is quadratic in the fluctuation fields. As a result the $3$- and $4$-point vertices involving $\hat{N}$ are generated by the EH-action. Analyzing the $N$-dependence of the kinetic and potential terms then shows that the kinetic term contributes to a $4$-point vertex coupling $\hat{N}\hat{N}h_{ij}h_{kl}$ which contains derivatives with respect to $\tau$ while there is no analogous contribution originating from $^{(3)}R$ which could provide the spatial derivatives required for making this vertex relativistic.} In the sequel, we will consider both projections of the flow simultaneously, referring to them as the $p_0$- and the $\vec{p}$-projection, respectively. Conceptually, the fact that these projections lead to slightly different (but qualitatively identical) results should be understood as tracing the RG flow of two different couplings which are distinguished in the non-relativistic setting while in the relativistic setup they are identified. 

\begin{table}[t!]
	\renewcommand{\arraystretch}{1.5}
		\begin{tabular}{ll}
			\hline\hline
			$p^1_\lambda(\lambda)$ & $12 \left(18 \lambda^4-116 \lambda^3+142 \lambda^2-61 \lambda+8\right)$\\
			$\tilde{p}^1_\lambda(\lambda)$ & $-27 \lambda^4+211 \lambda^3-259 \lambda^2+106 \lambda-12$\\
			\multirow{2}*{$p^2_\lambda(\lambda)$} & $10 \Big(68040 \lambda ^6+439308 \lambda ^5-1156914 \lambda ^4+1012717 \lambda^3$ \\ & $\qquad -419962 \lambda ^2+94584 \lambda -11760\Big)$ \\
			\multirow{2}*{$\tilde{p}^2_\lambda(\lambda)$} & $87480 \lambda ^6-1420428 \lambda ^5+2828930 \lambda ^4-2242429 \lambda ^3$ \\ & $+851098 \lambda ^2-166200 \lambda +17520$ \\ \hline
			\multirow{3}*{$p^3_\lambda(\lambda)$} & $408240 \lambda ^8-2262816 \lambda ^7+6128784 \lambda ^6-10347048 \lambda ^5$ \\
			&  $+10788945 \lambda ^4-6629544 \lambda ^3+2229588 \lambda ^2$ \\ & $-344176 \lambda +12704$ \\
			$\tilde{p}^3_\lambda(\lambda)$ & $4968 \lambda ^4-67980 \lambda ^3+94335 \lambda ^2-40380 \lambda +4148$ \\  \hline
			\multirow{2}*{$p^4_\lambda(\lambda)$} & $88248 \lambda ^5-180807 \lambda ^4+125287 \lambda ^3+3744 \lambda ^2$ \\ & $-30324 \lambda +6376$ \\
			$\tilde{p}^4_\lambda(\lambda)$ & $20055 \lambda ^4-32850 \lambda ^3+19676 \lambda ^2-771 \lambda -2158$  
			\\\hline \hline
		\end{tabular}
		\caption{\label{tab.1} Explicit form of the polynomials appearing in $\beta_\lambda$ given in eq.\ \eqref{betalambdaEH} (top block) and the functions $B_i(\lambda)$ of eq.\ \eqref{Bp0proj} (middle block) and eq.\ \eqref{Bpvecproj} (bottom block). The contributions arise from vertices without matter legs and agree with the pure gravity computation \cite{Saueressig:2023tfy}.}
\end{table}
The beta functions governing the $k$-dependence of \eqref{Gammaeffans} are computed along the lines detailed in \cite{Saueressig:2023tfy}. The result is conveniently expressed in terms of the dimensionless couplings
\begin{equation}
	g_k \equiv G_k k^2, \quad \lambda_k \equiv \Lambda_k k^{-2} \, , 
\end{equation}
together with the anomalous dimension $\eta_N \equiv (G_k)^{-1} \p_t G_k$. We find that
\be\label{betadef}
\p_t \lambda_k = \beta_\lambda(g_k,\lambda_k) \, , \qquad \p_t g_k = \beta_g(g_k,\lambda_k) \, ,   
\ee
where
\be\label{betalambdaEH}
\begin{split}
	\beta_g(g,\lambda) = & \, \left(2 + \eta_N \right) g \, , \\
	\beta_\lambda(g,\lambda) = & \, (\eta_N - 2) \lambda + \frac{g}{\pi}\bigg( \frac{p^1_\lambda(\lambda) + \eta_N \, \tilde{p}^1_\lambda(\lambda)}{24 \left(1-2 \lambda \right)^2 \left(2-3 \lambda\right)^2}  \\ & \, + \frac{p^2_\lambda(\lambda) + \eta_N \, \tilde{p}^2_\lambda(\lambda)}{7200 \left(1-2 \lambda \right)^3 \left(2-3 \lambda\right)^3} - \frac{N_s}{24\pi} \bigg) \, .
\end{split}
\ee
The anomalous dimension takes the form
\be\label{etaN}
\eta_N(g,\lambda) = \frac{g B_1(\lambda)}{1-g B_2(\lambda)} \, . 
\ee
The functions $B_1(\lambda)$ and $B_2(\lambda)$ obtained from projecting on the time-component of the external momentum ($p_0$-projection) read
\be\label{Bp0proj}
\begin{split}
	B_1^{p_0}(\lambda) = &\frac{48 \lambda ^2-60 \lambda +19}{2 \pi  (1-2 \lambda)^2 (2-3 \lambda )^2} \\ & -\frac{p^3_\lambda(\lambda)}{360 \pi  (1-2 \lambda)^4 (2-3 \lambda)^4}  + \frac{5 N_v}{6 \pi} + \frac{N_s}{12 \pi}\, ,\\
	B_2^{p_0}(\lambda) = & -\frac{48 \lambda ^2-60 \lambda +19}{12 \pi  (1-2 \lambda)^2 (2-3 \lambda)^2}  \\ & +\frac{\tilde{p}^3_\lambda(\lambda)}{720 \pi  (1-2 \lambda)^3 (2-3 \lambda)^2}\, . 
\end{split}
\ee
The projection onto the spatial momentum ($\vec{p}$-projection) contains an extra contribution
\be\label{Bpvecproj}
\begin{split}
	B_1^{\vec p}(\lambda) = & B_1^{p_0}(\lambda) -\frac{p^4_\lambda(\lambda)}{315 \pi  (2-3 \lambda )^2 (1-2 \lambda )^4} \, , \\
	B_2^{\vec p}(\lambda) = & B_2^{p_0}(\lambda) - \frac{\tilde{p}^4_\lambda(\lambda)}{630 \pi  (2-3 \lambda )^2 (1-2 \lambda)^3} \, .  \\	
\end{split}
\ee
The polynomials $p^i_\lambda$ and $\tilde{p}^i_\lambda$ capture the contributions of the vertices without matter legs and are tabulated in Table \ref{tab.1}. The contributions of the matter degrees of freedom appearing in $\beta_\lambda$ and $B_1$ are the main new computational result reported in this work.   

We close the derivation of the beta functions with two structural observations. Firstly, the matter fields give identical contributions to the flow, irrespective of whether we consider the $p_0$- or the $\vec{p}$-projection. This reflects the fact that the matter sector gives rise to propagators and vertices which are relativistic. Secondly, we observe that the vector degrees of freedom do not contribute in $\beta_\lambda$. This is owed to an exact cancellation of the vector contributions associated with the $3$- and $4$-point vertices. As a result $\beta_\lambda$ depends on $N_v$ through the anomalous dimension $\eta_N$ only.
%-------------------------------------------------------
\section{Fixed points and phase diagrams}
\label{sect.4}
%-------------------------------------------------------
The beta functions \eqref{betadef} constitute a coupled system of autonomous differential equations encoding the $k$-dependence of $g_k$ and $\lambda_k$. In order to understand the resulting phase diagram as a function of the number of matter fields, it is convenient to work along the following algorithm. We start with determining the singular loci of the beta functions. Inspecting $\beta_\lambda$, we find two lines where the beta function diverges
\be\label{singbl}
\begin{split}
\gamma_1^{\rm sing}(g) = (g, \lambda = 1/2) \, , \qquad \gamma_2^{\rm sing}(g) = (g, \lambda = 2/3) \, . 	
\end{split}
\ee
These divergences are generated by the propagators associated with the gravitational fields. In particular, $\gamma^{\rm sing}_2$ arises from the scalar sector comprising the fluctuation fields $\hat{N}$ and $\psi$. In addition, there is one singular locus associated with a divergence of $\eta_N$. This line appears when the denominator in eq.\ \eqref{Bpvecproj} vanishes and is thus conveniently parameterized as function of $\lambda$:
\be\label{etasing}
\gamma_3^{\rm sing}(\lambda) = (g = 1/B_2(\lambda), \lambda ) \, . 
\ee
The curve $\gamma_3^{\rm sing}(\lambda)$ is independent of the number of matter fields. Its exact position shows a mild dependence on the chosen projection scheme, which does not alter the phase diagram at a qualitative level though. RG trajectories cannot cross these singular lines. Solutions running into these loci then terminate there at a finite value of $k$. In addition, RG trajectories cannot cross the line
\be\label{g0divide}
\gamma_4(\lambda) = (g=0,\lambda) \, ,
\ee
since $\beta_g$ vanishes if $g=0$.

In order to understand the RG flow within the subspaces separated by these singular lines, it is convenient to proceed by finding the fixed points $(g_*,\lambda_*)$ of the system where, by definition,
\be\label{FPdef}
\beta_g(g_*,\lambda_*)=0 \, , \quad \beta_\lambda(g_*,\lambda_*)=0 \, . 
\ee
This system always exhibits the solution $(g_*,\lambda_*) = (0,0)$ to which we refer to as the Gaussian fixed point (GFP). In addition, the system possesses a number of non-trivial real roots associated with NGFPs. The properties of the RG flow in the vicinity of a fixed point are readily captured by the stability matrix
\begin{equation}
	{B^i}_j  := \frac{\partial \beta_{u^i}}{\partial u^j}\bigg|_{u=u_*} \, , \quad u^i = (g,\lambda) \, ,
\end{equation}
which arises from linearizing the beta functions at $u_*$. Defining the stability coefficients $\theta_I$ and right-eigenvectors $v_I^i$ through the relation ${B^i}_j v_I^j = - \theta_I v^i_I$, one readily verifies that RG trajectories along eigendirections with $\mathrm{Re}(\theta_I)>0$ are pulled into the FP for $k\rightarrow\infty$ and repelled from the FP for $k\rightarrow 0$, whereas RG trajectories along eigendirections with $\mathrm{Re}(\theta_I)<0$ are repelled from the FP for $k\rightarrow\infty$ and pulled into the FP for $k\rightarrow 0$. The former solutions then span the UV-critical hypersurface if the NGFP controls the UV behavior.

Inspecting the beta functions \eqref{betalambdaEH}, we first notice the existence of a GFP
\be
\text{GFP:} \qquad (g_*,\lambda_*) = (0,0) \, . 
\ee
This fixed point persists for all values $N_s$, $N_v$. Its critical exponents agree with classical power counting and do not receive quantum corrections 
\be
\theta_1^{\rm GFP} = 2 \, , \qquad \theta_2^{\rm GFP} = -2 \, . 
\ee

\begin{figure}
	\centering
	\includegraphics[width=0.8\linewidth]{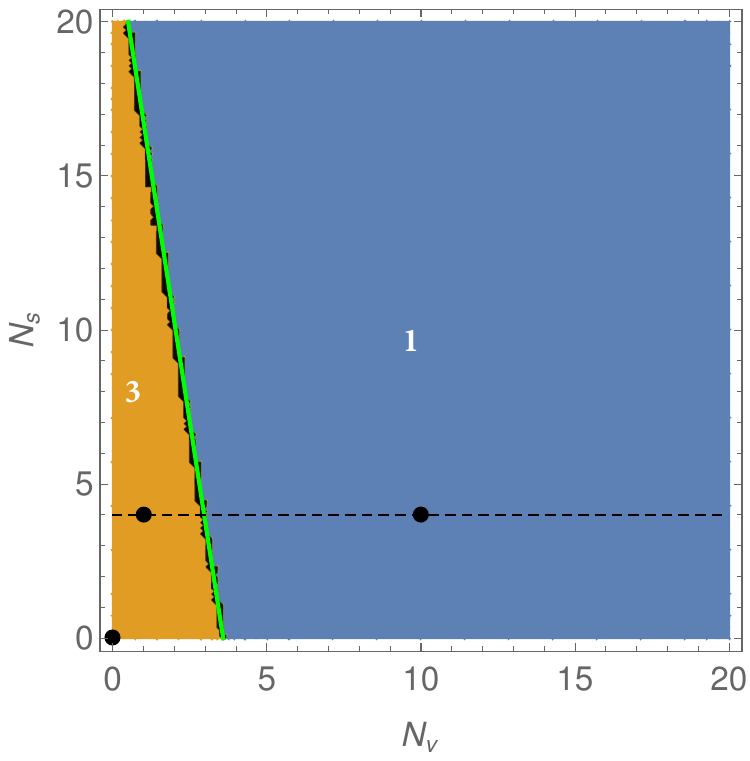} \\[1.5ex]
	\includegraphics[width=0.8\linewidth]{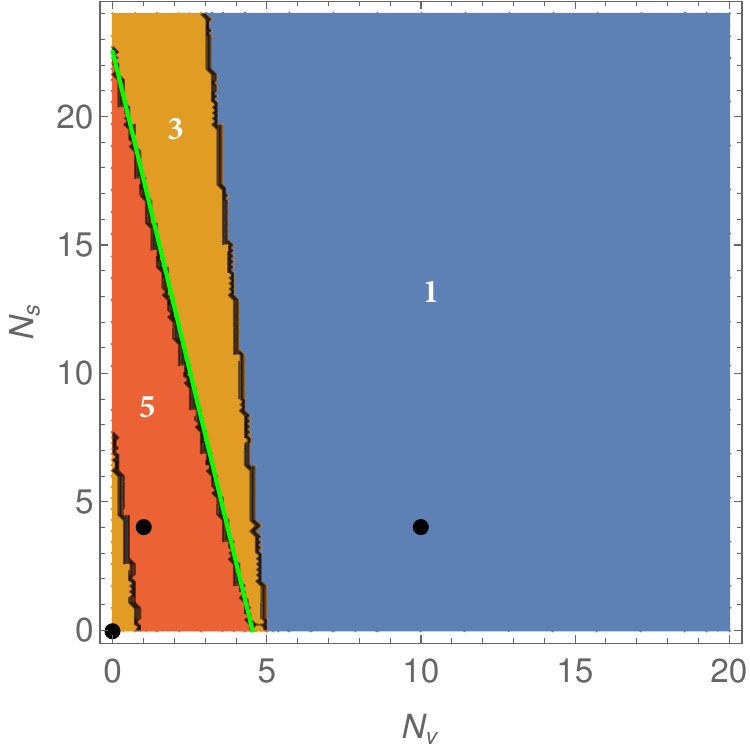}
	\caption{ Number of NGFPs as a function of the number of matter fields $(N_v,N_s)$ obtained within the $p_0$-projection (top panel) and $\vec{p}$-projection (bottom panel).
	Regions colored in blue, yellow, and orange host $1$, $3$, and $5$ NGFPs, respectively. The green lines are parameterized in eq.\ \eqref{lineparameterization} and mark the boundary where the NGFP relevant for the gravitational asymptotic safety program annihilates into the complex plane. The dashed line in the upper plot marks the fixed point structure shown in Fig.\ \ref{fig:fpcollision}.
}
\label{fig:NumberOfNGFPs}
\end{figure}

In addition, there are a number of non-trivial solutions of \eqref{FPdef}. Since the beta functions are polynomial in $g,\lambda$, their number is readily found by solving the system numerically. Varying the parameters $N_v, N_s$ deforms the root system. As a result, real roots may collide and vanish into the complex plane (fixed point annihilation). Similarly, new pairs of fixed points may emerge from the complex plane. Fig.\ \ref{fig:NumberOfNGFPs} summarizes the number of NGFPs found as a function of $N_s$, $N_v$. For pure gravity, $N_s=N_v=0$, we recover the 3 NGFPs identified in \cite{Saueressig:2023tfy}. The inclusion of the matter contributions leads to specific transition lines where the number of NGFPs changes by $2$. The lines
\be\label{lineparameterization}
\begin{split}
p_0-{\rm projection}: \qquad & N_s + 6.4 N_v = 23.1 \, ,  \\
\vec{p}-{\rm projection}: \qquad & N_s + 4.7N_v =  22.4 \, , 
\end{split}
\ee
turn out to be of special interest. They separate the regions with $3$ and $1$ NGFP ($p_0$-projection) and $5$ and $3$ NGFPs ($\vec{p}$-projection), respectively and are marked in green. At these lines the family of NGFPs (labeled by NGFP$_1$ in Table \ref{Tab.1}) collides with a second fixed point and ceases to exist.

Subsequently, we choose the three reference points in the $N_s$-$N_v$ plane. These correspond to the pure gravity case $N_s=N_v=0$, a reference system exhibiting a NGFP suitable for asymptotic safety, $N_s = 4, N_v= 1$, and $N_s = 4, N_v = 10$ where the NGFP$_1$ has been annihilated. The matter systems sit to the left and right of the separation line \eqref{lineparameterization} and are marked by the black dotes in Fig.\ \ref{fig:NumberOfNGFPs}. The complete information of the fixed point structure at these points is summarized in Table \ref{Tab.1}. A quick inspection shows that the NGFPs appear in distinct subspaces of the $g$-$\lambda$-plane which are separated by the lines \eqref{singbl}-\eqref{g0divide}. As a consequence, it is the NGFP which is the most interesting one from the perspective of the gravitational asymptotic safety program since it comes with $g_* > 0$ and is situated in the same subspace as the GFP.
\begin{table*}[!t]
\begin{center}
	\renewcommand{\arraystretch}{1.2}
\begin{tabular}{|c c| c | c | c  c  | c  c |  }
	\hline
	\multicolumn{2}{|c|}{matter fields} & projection & fixed point &\multicolumn{2}{c|}{position} & \multicolumn{2}{c|}{critical exponents} \\	
	\hline	
	$N_s$ & $N_v$ & ~ &~ & $g_*$ & $\lambda_*$ &  $\theta_1$ & $\theta_2$  \\ 
	\hline
	\multirow{6}{*}{$0$} & \multirow{6}{*}{$0$} & \multirow{3}{*}{$p_0$-projection} &NGFP$_1$ &  $1.43$  & $-0.12$          &  \multicolumn{2}{c |}{$4.42\pm1.38 i $} \\
	~&~&~&NGFP$_2$ & $-0.40$  & $0.14$  &  \multicolumn{2}{c |}{$1.41\pm3.84 i $} \\
	~&~&~&NGFP$_3$ & $5.86$  & $2.24$          & $9.37$  & $4.52$ \\
	\cline{3-8}
	~&~&\multirow{3}{*}{$\vec{p}$-projection}& NGFP$_1$ &  $0.78$  & $-0.07$          & $5.01$ & $2.73$ \\
~&	~&~& NGFP$_2$ &  $-0.23$ &   $0.22$  &  \multicolumn{2}{c |}{$-0.80\pm5.06 i $} \\
~&	~&~& NGFP$_3$ & $0.36$  & $1.04$          &$27.50$ &$-5.17$ \\
	\hline
	\multirow{8}{*}{$4$} & \multirow{8}{*}{$1$} & \multirow{3}{*}{$p_0$-projection} &NGFP$_1$ & $1.80$  & $-0.17$          &  \multicolumn{2}{c |}{$4.84\pm0.98 i $} \\
	~&~&~&NGFP$_2$ & $-0.39$  & $0.14$          & \multicolumn{2}{c |}{$1.51\pm3.66 i $} \\
	~&~&~&NGFP$_3$ &  $190.39$  & $-4.31$  &  \multicolumn{2}{c |}{$2.39\pm4.81 i $} \\
	\cline{3-8}
	& ~&\multirow{5}{*}{$\vec{p}$-projection}& NGFP$_1$ &  $0.96$  & $-0.09$          & $5.20$ & $2.59$ \\  
	~&~&~& NGFP$_2$ & $-0.23$  & $0.21$          & \multicolumn{2}{c |}{$-0.71\pm4.97 i $} \\
	~&~&~& NGFP$_3$ & $79.14$ &   $7.49$  & $12.18$ & $3.25$ \\
	~&~&~& NGFP$_4$ & $0.41$  & $1.06$          & $27.33$ & $-4.37$ \\
	~&~&~& NGFP$_5$ & $12.65$  & $-1.05$          &$13.88$ &$-6.93$ \\
	\hline
	\multirow{2}{*}{$4$} & \multirow{2}{*}{$10$} & \multirow{1}{*}{$p_0$-projection} & NGFP$_2$ & $-0.41$  & $0.10$  &  \multicolumn{2}{c |}{$1.77\pm2.35 i $}\\
	\cline{3-8}
	& ~&\multirow{1}{*}{$\vec{p}$-projection}& NGFP$_2$ &  $-0.27$ &   $0.20$  &  \multicolumn{2}{c |}{$-0.63\pm4.05 i $}\\
	\hline
\end{tabular}
\end{center}
\caption{\label{Tab.1} NGFPs found at the reference points marked by the black dots in Fig.\ \ref{fig:NumberOfNGFPs}. A NGFP suitable for asymptotic safety persists to the left of the separation line \eqref{lineparameterization} and is labelled by ``NGFP$_1$''. Families of other fixed points are then enumerated by subscripts.}
\end{table*}

\begin{figure}
	\centering
	\includegraphics[width=0.95\linewidth]{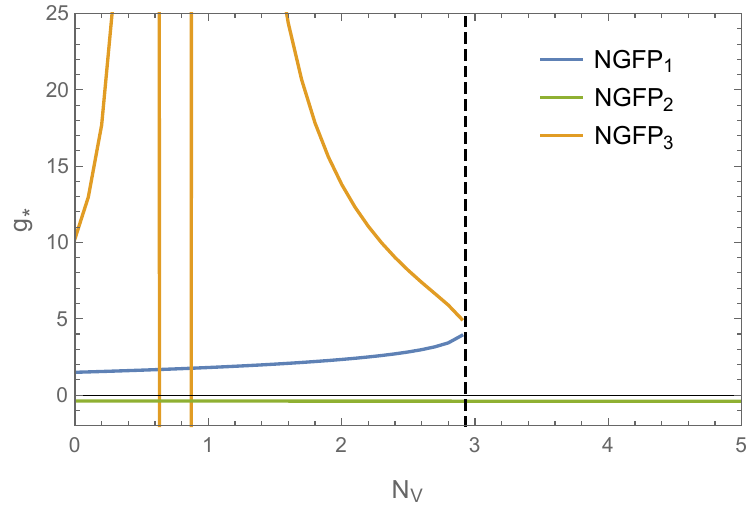}
	\caption{Illustration of the fixed point annihilation occurring at the boundaries \eqref{lineparameterization}. For concreteness, we picked the $p_0$-projection and fixed $N_s = 4$. The diagram gives the position $g_*$ of the NGFPs as a function of $N_v$ following a horizontal line in Fig.\ \ref{fig:NumberOfNGFPs}. The nomenclature follows the one for the reference system at $N_s=4$, $N_v=1$ given in Table \ref{Tab.1} which we use to label the families of NGFPs passing through this point. The family NGFP$_2$ (green line) is located at $g_* < 0$ and persists for all values $N_v$. Starting at $N_v=0$, the family NGFP$_3$ (orange line) first moves out to infinity, comes back in the region $g_* < 0$, and again crosses infinity to continue at $g_* > 0$. The two sign-changes in $g_*$ are marked by the vertical orange lines. The family NGFP$_3$ annihilates with the family NGFP$_1$ (blue line) at $N_v \simeq 2.95$, marked by the dashed vertical line.}
	\label{fig:fpcollision}
\end{figure}

The fixed point annihilation mechanism bounding the existence of the NGFP$_1$ as a function of $N_s$ and $N_v$ is illustrated in Fig.\ \ref{fig:fpcollision}.
For illustrative purposes, we fix $N_s = 4$ and display the position $g_*$ of the NGFP$_1$ (blue line) and NGFP$_3$ (dashed orange line) as a function of $N_s$. Following the lines towards increasing values $N_s$ we first observe that the NGFP$_3$ is shifted to infinity and taking a detour through the $g_* < 0$ region before reappearing from the top. This is reflected by the two vertical orange lines. Subsequently, the two fixed points collide at the black dashed vertical lines. At this point the family NGFP$_1$ ceases to exist. It is this mechanism which generates the bounds \eqref{lineparameterization}. 

\begin{figure*}[!p]
	\centering    
	\includegraphics[width=0.65 \columnwidth]{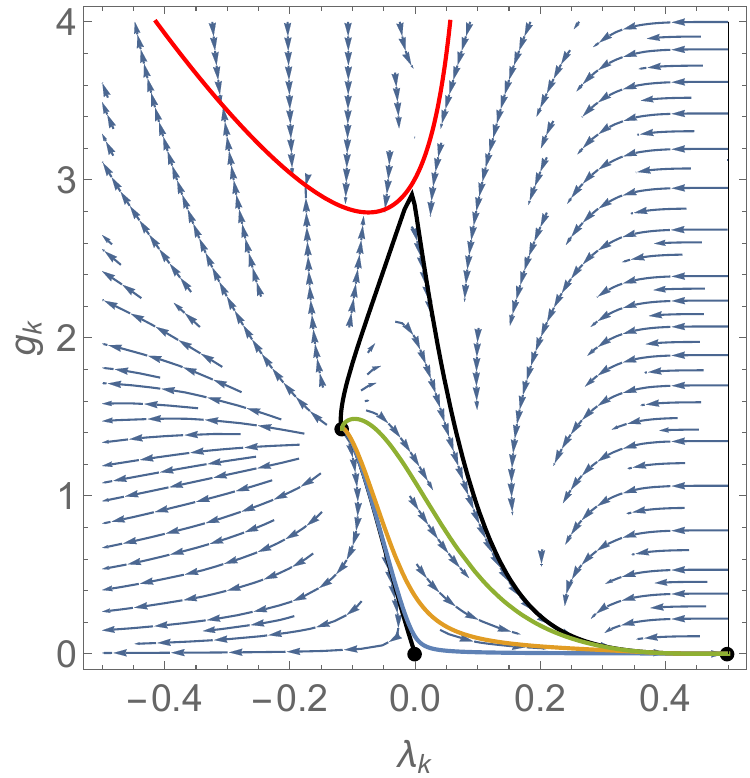} \;  \; \; \;
	\includegraphics[width=0.95 \columnwidth]{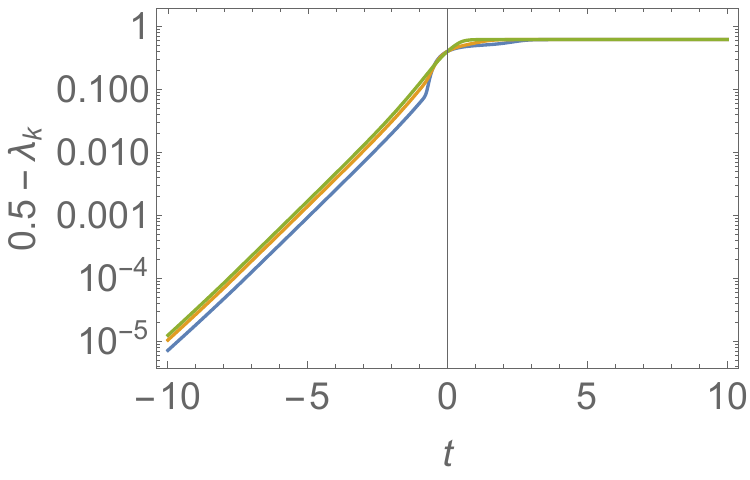} \\[2ex]
	\includegraphics[width=0.65 \columnwidth]{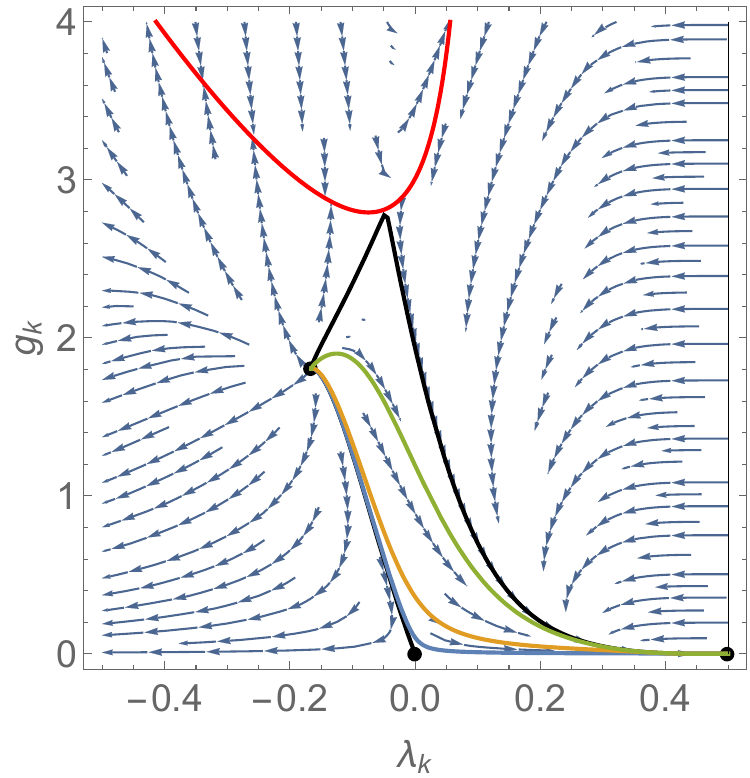} \; \; \; \;
	\includegraphics[width=0.95 \columnwidth]{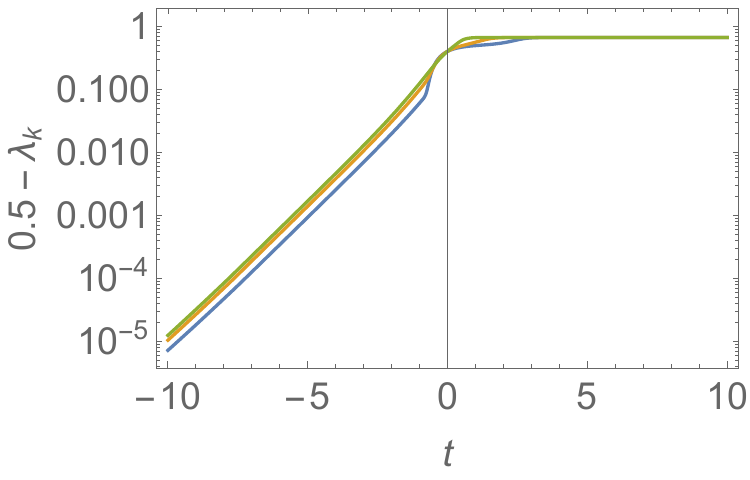} \\[2ex]
	\includegraphics[width=0.65 \columnwidth]{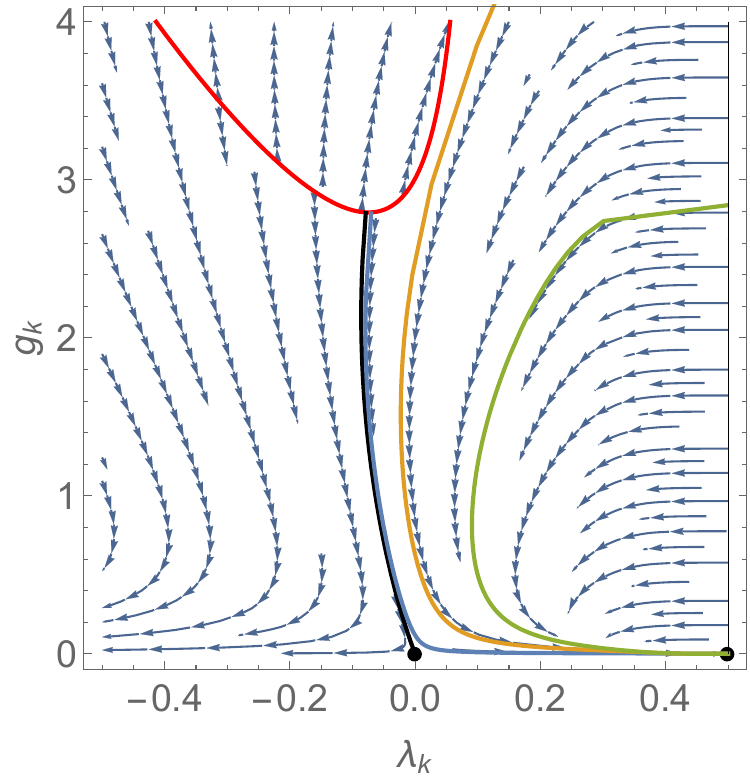} \;  \;\; \;
	\includegraphics[width=0.95 \columnwidth]{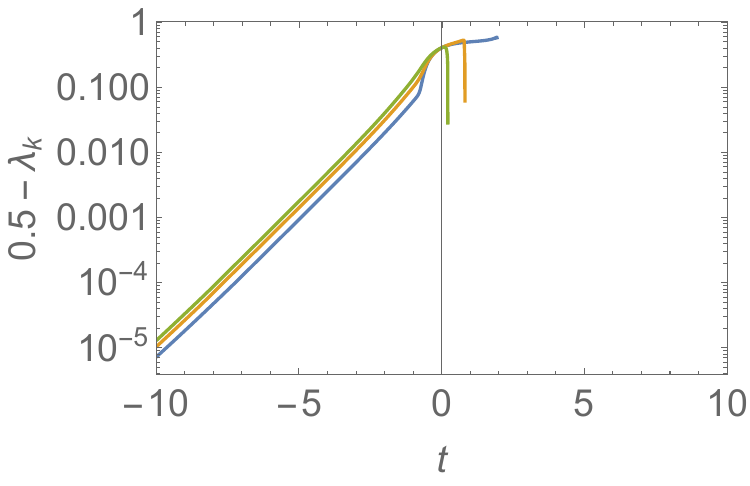}
	\caption{\label{phasetrans} Illustration of the typical phase diagrams (left column) resulting from the beta functions \eqref{betalambdaEH} obtained in the $p_0$-projection. From top to bottom, we show the cases, $N_s=N_v=0$ (pure gravity), $N_s=4, N_v=1$ (representing the gravity-matter systems featuring a fixed point within the NGFP$_1$ family), and $N_s = 4, N_v=10$ (representing the case where the NGFP$_1$ is absent). The red line and black line at the right display the singular loci $\gamma_1^{\rm sing}$ and $\gamma_3^{\rm sing}$, respectively, while the GFP, NGFP$_1$, and IR-FP are marked with black dotes. The black lines depict the RG trajectory ending at the GFP as $k \rightarrow \infty$. In addition the top and middle line contain the ``last'' trajectory connecting the NGFP$_1$ and IR-FP without terminating in the singular line $\gamma_3^{\rm sing}$. The region bounded by these black trajectories is populated by complete RG trajectories interpolating between the NGFP$_1$ for $k \rightarrow \infty$ and the IR-FP for $k \rightarrow 0$. The right column displays the flow of $\gamma_k$ as a function of the renormalization group time $t$ along the sample trajectories imprinted on the phase diagram (solid blue, orange, and green lines). For $t \gtrsim 5$ and $t \lesssim -5$ the flow is governed by the NGFP$_1$ and the scaling law \eqref{IR-FP-scaling}, respectively. When the NGFP$_1$ is absent (bottom row) the trajectories terminate at a finite value $t$ running either into the line $\gamma_1^{\rm sing}$ (green and orange line) or $\gamma_3^{\rm sing}$ (blue line).  All arrows point towards lower values of the coarse-graining scale $k$.}
\end{figure*}

At this stage we are in the position to discuss the phase diagram resulting from the beta functions \eqref{betalambdaEH}.  Our focus is on the region bounded by the lines $\gamma_1^{\rm sing}$ (right), $\gamma_3^{\rm sing}$ (top) and $\gamma_4$ (bottom). This region hosts the GFP and the NGFP$_1$, provided that the latter exists. The phase diagrams corresponding to the reference systems ($N_s = 0$, $N_v=0$), ($N_s = 4$, $N_v =1$), and ($N_s = 4$, $N_v=10$), whose fixed point structures are detailed in Table \ref{Tab.1}, are shown in the top line, middle line, and bottom line of Fig.\ \ref{phasetrans}, respectively. 
The left column then shows the phase portrait while the right column highlights the RG flow of $\lambda_k$ along the example RG trajectories highlighted in the phase portrait.

The phase diagrams originate from the interplay of the GFP, NGFP$_1$, and an IR-FP
\be\label{IRFP}
\text{IR-FP:} \qquad (g_*^{\rm IR}, \lambda_*^{\rm IR}) = (0, 1/2) \, ,
\ee
highlighted by the black dots. At the IR-FP, the beta functions are degenerate. The presence of the fixed point is then concluded based on the scaling behavior of the RG flow in its vicinity. Numerical integration of the beta functions in this region shows that
\be\label{IR-FP-scaling}
\text{IR-FP:} \quad (g_k - g_*^{\rm IR}) \sim c_1 e^{t} \, , \; (\lambda_*^{\rm IR} - \lambda_k) \sim c_2 e^{4 t} \, , 
\ee 
where $c_1$, $c_2$ $\in$ $\mathbb{R}$ specify the RG trajectory under consideration and $\sim$ is used to indicate that the scaling applies sufficiently close to the IR-FP. The scaling law eq.\ \eqref{IR-FP-scaling} is depicted in the right column of Fig.\ \ref{phasetrans}. 

The structure of the phase diagrams can be summarized as follows. Firstly, there is a single RG trajectory (black solid line) which ends at the GFP as $k=0$. If the NGFP$_1$ exists, this line emanates from it. Otherwise it emerges from the line $\gamma_3^{\rm sing}$. RG trajectories to the left of this line either terminate at $\gamma_3^{\rm sing}$ or reach the point $(g,\lambda) = (0,-\infty)$ in the limit $k=0$. The latter trajectories give rise to a finite and positive renormalized graviton mass $\mu^2 \equiv \lim_{k\rightarrow 0} -2 \Lambda_k > 0$. The region to the right of this separation line is separated into two subregions. In the presence of NGFP$_1$, trajectories within the black lines interpolate between this fixed point in the UV ($k \rightarrow \infty$) and the IR-FP in the IR ($k \rightarrow 0$). They are complete RG trajectories in the sense that they are well-defined for all values $k \in [0,\infty[$. They all lead to a vanishing renormalized graviton mass, $\mu^2 = 0$. This class of trajectories is absent if NGFP$_1$ has annihilated (see the bottom row). Finally, RG trajectories to the right of this region emanate from the singular lines $\gamma_1^{\rm sing}$ or $\gamma_3^{\rm sing}$. In the limit $k\rightarrow 0$ these are again captured by the IR-FP and thus result in $\mu^2 = 0$.

The most remarkable feature of the phase diagram is the IR-FP. This IR attractor ensures that $\mu^2 = 0$ for an entire class of RG trajectories. Thus there is no need to fine-tune the RG flow in order to achieve a zero renormalized graviton mass. This attractor mechanism exists for all values $N_s, N_v$. This property is easily argued based on the beta functions: the terms proportional to $N_s$ and $N_v$ come with a factor $g$. Thus the matter contributions decouple at this fixed point. This indicates that the IR-FP is a pure gravity effect: the scaling behavior \eqref{IR-FP-scaling} arises from the interplay of $g$ and $(1-2\lambda)$ vanishing at this point. This also implies that the IR-FP is an IR-effect which exists independently of the UV completion by the NGFP$_1$. Thus this feature should also be visible within an effective field theory approach to quantum gravity. 
%-------------------------------------------------------
\section{Conclusions and Outlook}
\label{sect.conclusions}
%-------------------------------------------------------
We studied the \emph{Wilsonian RG flow} of the graviton $2$-point function \eqref{Gammaeffans}. The closure of the flow equation underlying this analysis is provided by the foliated Einstein-Hilbert action supplemented by minimally coupled scalar and Abelian gauge fields as well as gauge-fixing and ghost contributions. We find that the NGFP associated with asymptotic safety of the pure-gravity system (found in \cite{Saueressig:2023tfy}) admits deformations by the matter degrees of freedom and persists unless the matter contributions exceed the bounds \eqref{lineparameterization}. At these lines, the fixed point family NGFP$_1$ annihilates with a second root of the beta functions and vanishes into the complex plane.

The most remarkable feature of the phase diagrams constructed in our work is the occurrence of an IR-FP which controls part of the phase space in the limit where all quantum fluctuations are integrated out. For fluctuation computations in the covariant setting, this feature was first observed in \cite{Christiansen:2014raa} while for foliated spacetimes it first appeared in the pure-gravity analysis \cite{Saueressig:2023tfy}. For RG trajectories attracted to this fixed point as $k \rightarrow 0$, the renormalized graviton mass is driven to zero dynamically. Notably, this fixed point is a pure gravity effect. It persists \emph{independently of the number of matter fields added to the system}. In combination with the bounds found for the existence of NGFP$_1$, this entails that the properties induced by the IR-FP \emph{are independent of a potential high-energy completion of the system} via the asymptotic safety mechanism. The IR-FP is a generic low-energy feature which should also be visible within an effective field theory approach to quantum gravity. The discovery of this feature is the main result of our work.

It is instructive to compare the findings of this work with complementary investigations available in the literature. RG flows of gravity-matter systems within the ADM formalism, evaluating the Wetterich equation at zeroth order of the fluctuation fields, have been analyzed in \cite{Biemans:2017zca}. The resulting phase diagrams are qualitatively similar to the ones shown in the top and middle row of Fig.\ \ref{phasetrans}. In this case, the new elements appearing at the fluctuation level are the IR-FP and the bounds on the number of matter fields cast by the annihilation of the NGFP-family. Both features are absent in the background computation \cite{Biemans:2017zca}. Complementary, the impact of matter fields on RG fixed points have also been studied in covariant fluctuation field computations \cite{Meibohm:2015twa,Christiansen:2017cxa,Eichhorn:2018akn,Eichhorn:2018nda,Pawlowski:2020qer}, also see \cite{Pawlowski:2023gym} for a very recent review. The scalar case studied in \cite{Eichhorn:2018akn} observed a NGFP getting shifted to infinity at around $N_s \approx 52$. For gauge bosons, \cite{Christiansen:2017cxa} showed the occurrence of the fixed point annihilation mechanism at $N_v \approx 13$. Our results on the gravity-matter fixed point structure (see Figs.\ \ref{fig:NumberOfNGFPs} and eq.\ \eqref{lineparameterization}) are in qualitative agreement with these findings, even though our computation, in general yields tighter bounds on the number of matter fields compatible with the NGFP-family. This may be a hint that more elaborate approximations, as the ones used in the covariant setting, weaken the bounds obtained in our work.

Determining the renormalized couplings entering in the $2$-point function \eqref{Gammaeffans} may constitute the first step in making a systematic connection between the gravitational asymptotic safety program and cosmological perturbation theory. This link motivates a number of generalizations. These include the generalization of the flat Euclidean background to Friedmann-Robertson-Walker spacetimes (extending the computations \cite{Biemans:2016rvp,Biemans:2017zca,Platania:2017djo} to the level of fluctuation fields) and the extension of the projection subspace by interactions associated with the (background) Hubble parameter. Furthermore, it is interesting to extend the fluctuation setting to also include the graviton $3$-point function, which carries information about non-Gaussianities. Ultimately, this strategy may allow to test asymptotic safety based on cosmological data (also see \cite{Chataignier:2023rkq} for a related discussion in the context of canonical quantum gravity). Encoding the gravitational dynamics in the ADM fields makes the link particularly transparent, as it allows to study the RG flow of physically relevant correlation functions directly.

An intricate property of our present computation is that the vertices involving fluctuations in the lapse function give rise to contributions which are not Lorentz-covariant. Essentially, the vertex structure treats the spatial and time-component of the external momentum on different footing. As a result, the projection of the RG flow onto the $2$-point function \eqref{Gammaeffans} leads to slightly different results when reading off the beta function of $G_k$ from the $p_0^2$- or the $\vec{p}^{\,2}$-component. This difference does not affect the structure of the phase diagram encoding the flow of \eqref{Gammaeffans} at a qualitative level though. Nevertheless, it would be desirable to compute the the Wilsonian RG flow of this $2$-point function in a Lorentz-covariant way. Since the gauge-fixing can be engineered in such a way that all $2$-point functions are Lorentz-covariant, it is tempting to try to supplement the gauge-fixing \eqref{eqn:gfaction} by terms which are non-linear in the fluctuation field. Since $h_{ij}$ is gauge-invariant, it is unlikely that such a procedure can generate the required contributions to the $hh\hat{N}\hat{N}$-vertex. An alternative is the strategy employed in \cite{Knorr:2018fdu}, which suggests to modify the linear split \eqref{backgrounddecomposition} by adding terms which are non-linear in the fluctuation field.\footnote{In a covariant setting, a prominent example for a non-linear relation between $g_{\mu\nu}$ and $h_{\mu\nu}$ is the exponential split $g_{\mu\nu} = \gb_{\mu\alpha} [e^h]^\alpha{}_\nu$ discussed in detail in \cite{Ohta:2016npm,Ohta:2016jvw}.} This question certainly warrants further investigation and we hope to come back to it in the future.

%-------------------------------------------------------

%-------------------------------------------------------
\section*{Acknowledgments}
We thank J.\ Ambj{\o}rn, T.\ Borck, T.\ Budd, M.\ Becker, R.\ Loll, D.\ Nemeth, and C.\ Wetterich for discussion. JW acknowledges the China Scholarship Council (CSC) for
financial support.
%-------------------------------------------------------

\appendix
	%----------------------------------------------
\section{The ghost sector}
\label{App.ghosts}
  %------------------------------------------------
We complete the ansatz for $\Gamma_k$ providing the closure of the projected Wetterich equation by giving the explicit form of the ghost action. Evaluating the standard formula for the Fadeev-Popov procedure for the gauge-fixing functionals \eqref{eqn:gffunctionals} gives for $F_i$  
  \begin{equation}\label{Sghost}
  	\begin{split}
  		\Gamma^{\text{vec}}_{\text{ghost}}  =& \int d\tau d^3y  ~\bar{b}^i \Big[\partial^2_\tau N_i c + \partial_\tau \sigma^{kl} N_k N_l \partial_i c +  \partial_\tau N^2 \partial_i c \\ & -   \partial_i \partial_\tau N c 
  		- \partial_i N_m \partial^m c + \partial_i N N_k \sigma^{kl} \partial_l c +   \partial^j N_i \partial_j c \\ & +   \partial^j N_j \partial_i c  
  		 - \frac{1}{2}   \partial_i  c \partial_\tau \bar{\sigma}^{mn}\sigma_{mn}
  		  +  ~  \partial^j c \partial_\tau \sigma_{ij} \\ & 
  		+  \partial_\tau b^k \partial_k N_i +   \partial_\tau N_k \partial_i b^k +  ~  \partial_\tau \sigma_{ki} \partial_\tau b^k  -   \partial_i b^k \partial_k N \\
  		&- \frac{1}{2}  ~ \partial_i  b^k \partial_k \bar{\sigma}^{mn}\sigma_{mn}-  ~ \partial_i \sigma_{jk} \partial^j b^k  +    \partial^j b^k \partial_k \sigma_{ij}\\
  		& +  ~  \partial^j \sigma_{jk} \partial_i b^k+   \partial^j \sigma_{ik} \partial_j b^k \Big].
  	\end{split}
  \end{equation}
  The analogous computation for $F$ yields
  \begin{equation}\label{Vghost}
  	\begin{split}
  		\Gamma^{\text{scalar}}_{\text{ghost}}  =& \int d\tau d^3 y   ~\bar{c} [ \partial_\tau^2 c N + ~ \partial_\tau b^k \partial_k N - ~\partial_\tau N N_j \sigma^{ij}\partial_i c \\ &
  		+ ~ \partial^i \partial_\tau N_i c + ~ \partial^i b^k \partial_k N_i + ~  \partial^i N_k \partial_i b^k + ~ \partial^i \sigma_{ki} \partial_\tau b^k 
  		\\ &
  		- \partial_\tau \sigma_{ik}\partial^i b^k+ ~ \partial^i \sigma^{kl} N_k N_l \partial_i c + ~ \partial^i N^2 \partial_i c \\
  		& - \frac{1}{{2}} ~ \partial_\tau c \partial_\tau  \bar{\sigma}^{ij}\sigma_{ij}- \frac{1}{{2}}~ \partial_\tau b^k \partial_k \bar{ \sigma}^{ij}\sigma_{ij}- ~ \partial_\tau N^i \partial_i c]
  	\end{split}
  \end{equation}
  Here all the derivatives act on the right, $\partial_\tau N_i c =N_i (\partial_\tau c) +(\partial_\tau N_i) c$, in order to reduce the length of the formula. Furthermore, all indices are raised and lowered with the flat background metric $\delta_{ij}$.
  %------------------------------------------------
%\end{appendices}

%----------------------------------------------------------
\bibliographystyle{elsarticle-num} 
%----------------------------------------------------------

%----------------------------------------------------------
\end{document}